\documentclass{article}
\usepackage{spconf,amsmath,graphicx}
\usepackage{subcaption}
\usepackage{amsmath,amsfonts,graphicx}

\usepackage[T1]{fontenc} 
\usepackage[utf8]{inputenc} 
\usepackage{graphicx}
\usepackage{color}
\usepackage{wasysym}
\usepackage{graphicx}
\usepackage{textcomp}
\usepackage{xcolor}
\usepackage{tabularx,adjustbox,booktabs}
\usepackage{colortbl}
\usepackage{multirow}
\usepackage[font=small,labelfont=bf]{caption}
\usepackage{subcaption}
\usepackage{mathtools}
\usepackage{makecell}
\usepackage{lineno}
\nolinenumbers

\title{perceptual musical features for interpretable audio tagging}
\name{
Vassilis Lyberatos, Spyridon Kantarelis, Edmund Dervakos and Giorgos Stamou
}
\address{
National Technical University of Athens, Athens, Greece
}

\begin{document}
\small
%
\maketitle
\begin{abstract}
In the age of music streaming platforms, the task of automatically tagging music audio has garnered significant attention, driving researchers to devise methods aimed at enhancing performance metrics on standard datasets. Most recent approaches rely on deep neural networks, which, despite their impressive performance, possess opacity, making it challenging to elucidate their output for a given input. While the issue of interpretability has been emphasized in other fields like medicine, it has not received attention in music-related tasks. In this study, we explored the relevance of interpretability in the context of automatic music tagging. We constructed a workflow that incorporates three different information extraction techniques: a) leveraging symbolic knowledge, b) utilizing auxiliary deep neural networks, and c) employing signal processing to extract perceptual features from audio files. These features were subsequently used to train an interpretable machine-learning model for tag prediction. We conducted experiments on two datasets, namely the MTG-Jamendo dataset and the GTZAN dataset. Our method surpassed the performance of baseline models in both tasks and, in certain instances, demonstrated competitiveness with the current state-of-the-art. We conclude that there are use cases where the deterioration in performance is outweighed by the value of interpretability.

\end{abstract}

\section{Introduction}\label{sec:introduction}

Music audio tagging encompasses a range of tasks centered on assigning descriptive tags to music tracks, offering particular utility in the context of organizing extensive music collections. These tags encompass musically relevant information such as instruments \cite{eronen2001comparison}, genres \cite{dervakos2021genre}, mood \cite{trohidis2011multi}, emotions \cite{soleymani20131000,lyberatos2023employing}, harmony \cite{chen2019harmony}, rhythm \cite{takeda2004rhythm}, and additional metadata \cite{lisena2022midi2vec}. The majority of methods employed to address these tasks involve feature extraction and machine learning (ML). However, there has been a shift towards minimizing the feature extraction step, often limited to acquiring only Mel-frequency cepstrum coefficients (MFCCs), with a growing inclination towards end-to-end deep learning models \cite{pons2018end}. This trend towards deep learning models brings with it increased opacity and challenges in providing explanations, particularly in scenarios where the ground truth itself can be subjective and open to interpretation \cite{buisson2022ambiguity,koops2019annotator}, as is the case with mood tags.

In these cases, the reliability of the training data dictates the reliability of the model, and without interpretability, it is difficult to detect flaws or biases in either data or models \cite{Rudin2019}. For music especially, where real-world data tend to be diverse, contrary to available datasets \cite{holzapfel2018ethical}, it is essential to be able to detect and fix biases to make robust systems that can be applied in real-world settings \cite{mishra2017local}. As an extreme example, consider a mood tagging system that was trained only on pop and rock music. We cannot expect such a system to be generalized for all music genres, and it would not be possible to predict how it would behave if it were fed for instance classical music. Furthermore, we cannot expect an end-user to know the distribution of the training dataset, and the nuances of training ML models, thus such a system would not be applicable in practice, regardless of how well it performed on a test set. Contrarily, if the mood tagging system itself were interpretable, it would be easier for a user to understand why the system assigns specific tags to samples, making it more trustworthy and more likely to be applied in a real-world setting.

\begin{figure}[!t]
    \large
    \centering
    \includegraphics[scale=0.36]{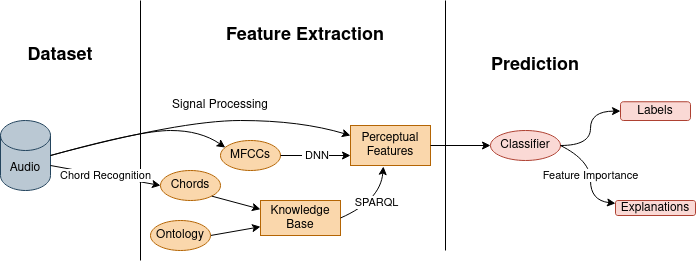}
    \caption{Overview of the proposed pipeline}
    \label{fig:sysarch}
\end{figure}


In recent years, efforts have emerged aimed at addressing the interpretability challenge within music audio tagging systems \cite{won2019toward}. Perceptual musical features are used that are readily comprehensible to human users~\cite{friberg2014using,aljanaki2018data,chowdhury2019towards,chowdhury2021towards,chowdhury2021tracing}. These features encompass characteristics imbued with both musical and acoustic significance, such as tonality, rhythmic stability, or loudness, and they exhibit a strong correlation with emotions \cite{gabrielsson2010role,wedin1972multidimensional}. Due to this inherent correlation, these features have demonstrated their efficacy in Music Emotion Recognition (MER) tasks. These characteristics are rooted in musical concepts and theoretical knowledge, as noted in a survey \cite{han2022survey}, though quantitative studies on them remain relatively scarce. In our research, we build upon these concepts and execute a structured approach for achieving interpretable music audio tagging (Fig.~\ref{fig:sysarch}), harnessing the power of perceptual features. Our specific contributions can be succinctly summarized as follows:

\begin{itemize}
    \item In our study, we integrate three distinct methods for extracting perceptual musical features: signal processing, deep neural networks, and symbolic knowledge. Particularly noteworthy is our introduction of a novel set of features directly tied to music harmony. We demonstrate that these information sources are mutually complementary, significantly enhancing classifier performance across two datasets.
    \item We interpret model predictions, using four different methodologies, SHAP \cite{SHAP}, XGBoost \cite{xgboost} feature importance, permutation-based feature importance, and an ablation study.
    \item Utilizing the interpretability methods, we discuss how we could alleviate issues with ML-based audio tagging, such as detecting biases of the training dataset which would be obscured in a deep learning setup, and better comprehend ambiguous labels.
\end{itemize}


\section{Extracting perceptual features}\label{sec:features}


\subsection{Datasets}\label{datasets}

We employed two datasets in our experiments, the MTG-Jamendo dataset~\cite{mtg} and the GTZAN dataset~\cite{tzanetakis2002musical}. The MTG-Jamendo dataset serves as an open resource for music audio tagging. This dataset was constructed using music accessible through the Jamendo platform, offered under Creative Commons licenses, and enriched with tags provided by content uploaders. Boasting a substantial scale, it encompasses more than 18,486 complete audio tracks, that we used, annotated with 56 tags that encompass various mood and theme categories. Our choice of this specific dataset was predicated on the anticipation that interpretability would offer heightened value in a task involving partially subjective aspects, such as mood detection~\cite{kim2010music}.

The GTZAN dataset~\cite{tzanetakis2002musical} is a collection comprising 1000 audio tracks of 30 seconds and features 10 genres, with 100 tracks representing each. The audio files are all 16-bit monophonic recordings in .au format, sampled at 22kHz. We opted to use this dataset to assess the generalizability of our proposed pipeline on a smaller scale and a different corpus.

\subsection{Methodologies}

To extract perceptual features from the datasets, we employed three different domains: Symbolic Knowledge, Deep Neural Networks, and Signal Processing.

By utilizing \textbf{symbolic knowledge}, we created a new set of interpretable features that derive from music theory. To accomplish this, we constructed a knowledge base to extract music features from each track's chords. We selected Omnizart \cite{wu2021omnizart} from among many chord recognition models, based on its reported performance. Our knowledge base consists of chords as instances, and classes and subclasses from the Functional Harmony Ontology (FHO)~\cite{KANTARELIS2023100754}. These classes and subclasses describe the harmonic function of each chord in a music track. Using SPARQL queries, we retrieved the instances of certain classes to create useful features, such as the frequency of specific harmonic patterns (e.g., \textit{dominant-dominant-tonic}), in the form of n-grams. In total, we created a 32-dimensional feature vector $x_{h} \in{\mathbb{R}^{32}}$. First, we created \textit{Dominants} and \textit{Subdominants} by calculating the ratio between the sum of dominant and subdominant retrieved chords and the total number of chords, respectively. Additionally, by defining the function of a chord in any chord progression, we developed 30 more features based on bigrams and trigrams (e.g., \textit{sub-dom}, \textit{sub-sub-dom}). These features were defined by calculating the ratio of a specific bigram or trigram to the total number of bigrams or trigrams in the analysis, respectively. To construct these features, we also designated some chords as tonics - chords that follow a dominant without serving another function, and globs - major dominant chords that produce a great deal of harmonic tension.

We leveraged the power of \textbf{deep neural networks} (DNNs) to efficiently capture perceptual features from a dataset curated by music experts~\cite{aljanaki2018data}. To build our DNN model, we used a vgg-ish architecture trained with multiple linear regression. We first extracted MFCCs with 40 coefficients, utilizing a window and FFT size of 2048, from the initial 15 seconds of each audio file. This procedure resulted in an input MFCC \begin{math} K \in \mathbb{R} ^{S \times P} \end{math}, where $S$ is the number of segments and $P$ is the number of MFCCs. The model's output is a feature vector $x_{m}\in \mathbb{R}^p$, where \begin{math} p\in \mathbb{N}\end{math} corresponds to the number of extracted features. We trained our model on the "Mid Level Perceptual Music Features" dataset~\cite{aljanaki2018data}, comprising 5000 songs, each lasting 15 seconds in audio format. Each song is annotated by music experts with seven scalar perceptual features: \textit{Melodiousness}, \textit{Articulation}, \textit{Rhythmic Stability}, \textit{Rhythmic Complexity}, \textit{Dissonance}, \textit{Tonal Stability}, and \textit{Minorness}. 

Utilizing \textbf{signal processing} we extracted acoustically meaningful features. In order to choose features appropriate for our task a filtering process is essential having as a criterion their interpretability. If $x$ is the input audio file, we define our process as a function $E$, where \begin{math} E(x)=x_{s}\end{math}, \begin{math}x_{s} \in \mathbb{R}^q\end{math} and \begin{math} q \in \mathbb{N}\end{math} ($q$ the number of the extracted features). For this approach, we used features extracted with a signal processing tool called Essentia \cite{bogdanov2013essentia}. The features were filtered out from three music experts, retaining only those that hold acoustical and musical significance and are easily understandable by humans. They ended up choosing 23 scalar features\footnote{https://essentia.upf.edu/streaming\_extractor\_music.html\#music-descriptors}: \textit{Danceability}, \textit{Loudness}, \textit{Chords Changes Rate}, \textit{Dynamic Complexity}, \textit{Zero Crossing Rate}, \textit{Chords Number Rate}, \textit{Pitch Salience}, \textit{Spectral Centroid}, \textit{Spectral Complexity}, \textit{Spectral Decrease}, \textit{Spectral Energyband High}, \textit{Spectral Energyband-Low}, \textit{Spectral Energyband Middle High}, \textit{Spectral Energyband Middle Low}, \textit{Spectral Entropy}, \textit{Spectral Flux}, \textit{Spectral Rolloff}, \textit{Spectral Spread}, \textit{Onset Rate}, \textit{Length}, \textit{BPM}, \textit{Beats Loud} and \textit{Vocal-Instrumental}.

\textit{The description of the features and the code for the implementation of the experiments can be found in the GitHub repository~\footnote{
https://github.com/vaslyb/PerceptibleMusicTagging
}}. By concatenating $x_{h}, x_{m}, x_{s}$, and normalizing the values with a standard scaler, we end up with the feature vector used to train the models.

\begin{figure}[!b]
    \centering
    \begin{subfigure}{0.45\textwidth}
        \includegraphics[width=\linewidth]{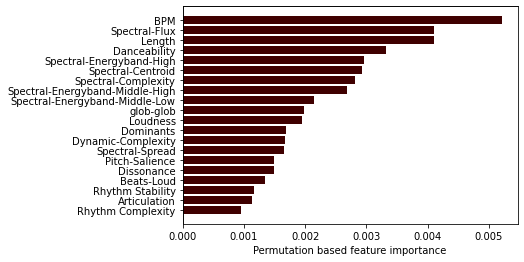}
        \label{fig:66}
    \end{subfigure}\hfil 
    \begin{subfigure}{0.45\textwidth}
        \includegraphics[width=\linewidth]{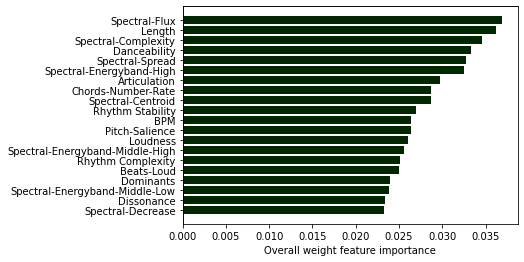}
        \label{fig:777}
    \end{subfigure}\hfil 
    \caption{Overall and permutation-based feature importance of the model trained on the MTG Jamendo dataset}
    \label{fig:permutation}
\end{figure}


\section{Experiments}\label{sec:experiments}



When dealing with a set of perceptual features, there exist two strategies for achieving explainable predictions of audio tags. The first approach involves employing explainable-by-design ML models
, like Decision Trees. The second approach revolves around the utilization of post hoc black-box explanation techniques capable of elucidating the functioning of any ML model \cite{guidotti2018survey}. In our research, we adopt a combination of both approaches. We incorporate explainable-by-design models due to their capacity to provide readily understandable outputs when fed with human-interpretable features as input. Additionally, we leverage post hoc black-box explanation methods, such as SHAP values, to enhance our interpretability efforts. 


\subsection{Classification}


After experimenting with a wide array of traditional ML classifiers, we ended up choosing XGBoost as the most effective classifier. XGBoost is an ensemble ML classifier based on decision trees that employs a gradient-boosting framework. Our model achieved a ROC-AUC score of 0.729, in the multi-label mood recognition task of the MTG Jamendo dataset, slightly surpassing the baseline of 0.725~\cite{tovstogan2021media}
while falling slightly below the performance of state-of-the-art intelligent systems 0.769~\cite{bour2021media}.
Similar results were obtained when applying XGBoost to the multi-class classification task using the GTZAN dataset, where we achieved an accuracy of 0.79, showing a slight deviation from the state-of-the-art performance 0.84~\cite{gtzan1,gtzan2}. While our findings may not outperform the state-of-the-art deep learning models, the presence of interpretable model outputs serves the valuable purpose of identifying potential biases within the model. Furthermore, we achieve commendable results using models characterized by a limited number of trainable parameters. 

\subsection{Explanations}

Interpretability played a pivotal role in the selection of XGBoost as the predictive modeling technique. It enabled us to gain insights into the significance of input features in the overall predictions. This characteristic has rendered it a valuable tool for improving our understanding of the classification process. Our study utilized four methods to extract feature importance from our XGBoost model, including the built-in feature importance, permutation method, SHAP values, and ablation studies.

Initially, we utilized the \textbf{XGBoost feature importance}, which involved assessing the weight of features based on their frequency in trees. By aggregating this information across all tags, we obtained a holistic overview of the features crucial to the task. Additionally, examining the weight of each specific tag allowed us to discern the importance of features on a tag-by-tag basis.

To further enhance our insights, we employed the \textbf{permutation method} to compute a second set of feature importance explanations. This involved randomly shuffling the values of each feature and observing the resulting degradation of the model’s score. This approach provided an overall perspective on the importance of features across all mood tags.

Subsequently, we incorporated \textbf{SHAP values} to extract feature importance, leveraging game theory concepts to estimate the contribution of each feature to predictions. Unlike the weight and permutation methods, SHAP values indicated whether features impacted the predictions positively, or negatively. The results were visualized through plots, showcasing the distribution of impacts each feature had on the model output, sorted by the sum of SHAP value magnitudes over all samples.

In line with recommended practices from the literature, we conducted an \textbf{ablation study} with different feature groups, categorized as Mid-level, Harmonic, and Signal based on their origin. By experimenting with all possible combinations of these groups, we gained valuable insights into the distinct impacts each set of features had on the model's performance.

\begin{figure}[!t]
    \centering
    \begin{subfigure}{0.37\textwidth}
        \includegraphics[width=\linewidth]{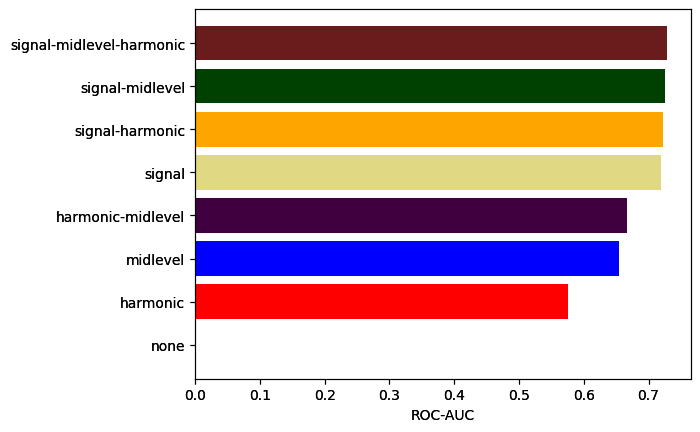}
        \label{fig:66666}
    \caption{MTG Jamendo}
    \end{subfigure}\hfil 
    
    \begin{subfigure}{0.37\textwidth}
        \includegraphics[width=\linewidth]{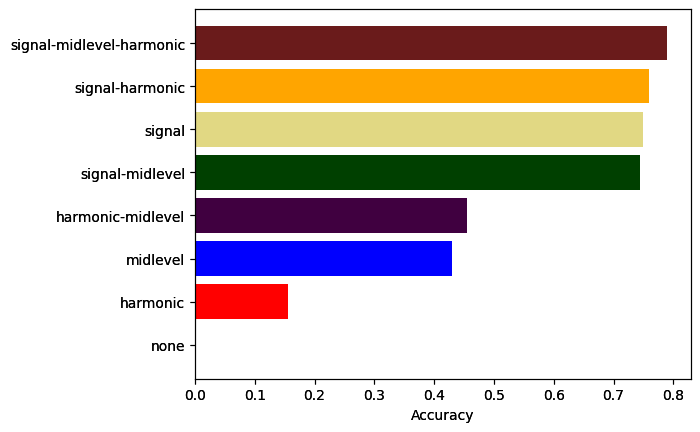}
        \label{fig:7}
    \caption{GTZAN}
    \end{subfigure}\hfil 
  \caption{Ablation studies with different three groups of features: harmonic, signal, and midlevel on both datasets}
  \label{fig:both}
\end{figure}


\begin{figure}[!b]
    \centering

    \begin{subfigure}{0.41\textwidth}
        \includegraphics[width=\linewidth]{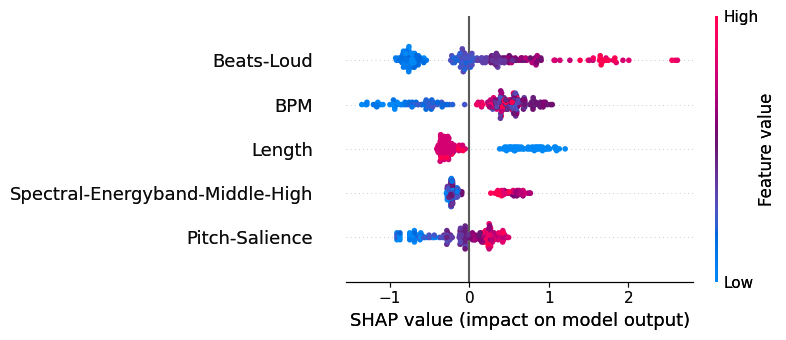}
        \label{fig:666}
    \end{subfigure}\hfil 
    \begin{subfigure}{0.41\textwidth}
        \includegraphics[width=\linewidth]{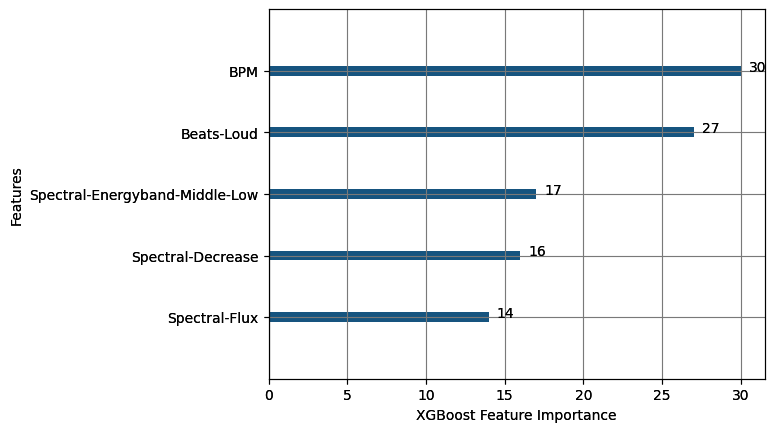}
        \label{fig:77}
    \end{subfigure}\hfil 



    \small
    \caption{Interpretations of the prediction for the label \textit{Disco} in the GTZAN dataset based on SHAP values and model's feature importance.}
\label{fig:gtzan}
\end{figure}


\section{Discussion}\label{sec:dicussion}

The results of permutation-based and the overall weight XGBoost feature importance for the Jamendo dataset are shown in Fig.~\ref{fig:permutation}. The two methods produce similar results, which indicates that the displayed features are likely indeed important. Signal processing features were the most important for predictions, notably 
\textit{Spectral-Flux}, \textit{Spectral-Complexity} and \textit{Spectral-Centroid}.

We obtained similar results in our ablation study (Fig.~\ref{fig:both}), where we found that signal processing features were the most useful, while harmonic features were the least useful for both datasets. Considering that harmonic features are extracted from the chord progression, this observation could be attributed to the quality of the chord recognition tool. 
Harmonic features had a substantial impact only on predicting the \textit{Blues} labels, a genre where chord progressions are distinct and easy to detect throughout a track.
In terms of mid-level features, they appeared to be significant in improving the ROC-AUC metric for the Jamendo dataset. For both datasets, the best results were achieved by the combination of all features.

\begin{figure}[]
    \centering

    \begin{subfigure}{0.41\textwidth}
        \includegraphics[width=\linewidth]{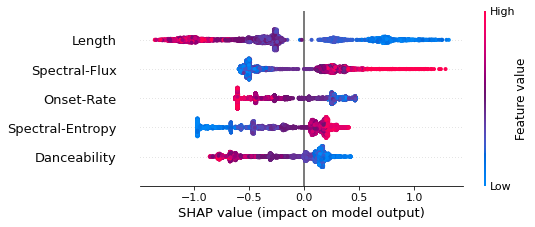}
        \label{fig:6666}
    \end{subfigure}\hfil 
    \begin{subfigure}{0.41\textwidth}
        \includegraphics[width=\linewidth]{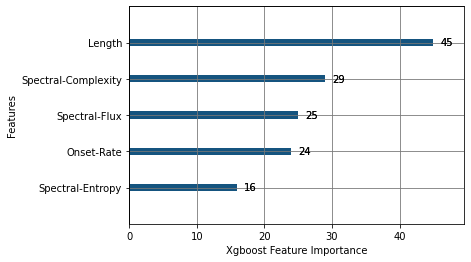}
        \label{fig:7777}
    \end{subfigure}\hfil 



    
\caption{Interpretations of the prediction for the label \textit{Trailer} in the MTG Jamendo dataset based on SHAP values and model's feature importance}
\label{fig:trailer}
\end{figure}


We show examples of interpreting feature importance computed with SHAP values and XGBoost for predicting mood labels in the MTG Jamendo dataset. Specifically, we analyze one of the best-performing labels (ROC-AUC > 0.8) as shown in Fig.~\ref{fig:trailer}. For the highly-predictable
\textit{Trailer} mood predictions rely on low values of \textit{Length} and high values of \textit{Spectral-Flux}, a feature that is used for voice detection. This is aligned with the fact that trailers are usually shorter than songs and have voice-overs.


For GTZAN, we show examples of the interpretation for one predicted genre, \textit{Disco} that reached 0.72 F1-score (see Fig.~\ref{fig:gtzan}). For this label, it is noticeable that high values of \textit{Beat-Loud} and middle values of \textit{BPM} are the most prevalent features. These observations align with the common perception of Disco music as a genre characterized by distinguishable beats, easy to dance to.


The above examples are meant to illustrate the value of interpretability for the audio tagging task. We can gain insights on a case-by-case basis which would not be possible in an end-to-end setup. Whether or not these insights are worth the additional work needed for feature extraction, and the slight deterioration in performance, depends on each application's priorities. We argue that interpretability should be a priority for music audio tagging systems, as they suffer from label ambiguity~\cite{gomez2020joyful}, in addition to being at risk of the typical biases of MIR datasets~\cite{mckay2006musical}.

\section{Conclusions and future work}\label{sec:conclusion}

We have created a specialized pipeline for music audio tagging with an emphasis on delivering interpretable results. Our approach incorporates established feature extraction methods alongside the introduction of a fresh set of features derived from the realm of music harmony concepts. Our experimental findings underscore the pivotal role of feature engineering within our proposed pipeline, as it exerts an influence on both performance and interpretability. It is imperative that each feature's meaning remains comprehensible to humans, emphasizing the significance of this step.

For future work, we plan to further investigate ways for improving performance given the explanations of predictions. 
Furthermore, we plan to study explanations regarding not only their performance improvement capabilities but also their usefulness and informativeness for end-users of music audio tagging systems. 
Finally, we plan to extend the feature extraction procedure with more features that are based on notions of music theory.

\section*{Acknowledgments}

We express our sincere gratitude to Christos Garoufis for his useful insights and thoughtful comments on our work. The research project is implemented in the framework of H.F.R.I call “Basic research Financing (Horizontal support of all Sciences)” under the National Recovery and Resilience Plan “Greece 2.0” funded by the European Union –NextGenerationEU(H.F.R.I. Project Number: 15111 - Emotional Artificial Intelligence in Music Expression).

\bibliographystyle{IEEEbib}
{\footnotesize
\bibliography{strings,refs}}

\begin{thebibliography}{10}

\bibitem{eronen2001comparison}
Antti Eronen,
\newblock ``Comparison of features for musical instrument recognition,''
\newblock in {\em WASPAA}. IEEE, 2001.

\bibitem{dervakos2021genre}
Edmund Dervakos, Natalia Kotsani, and Giorgos Stamou,
\newblock ``Genre recognition from symbolic music with cnns,''
\newblock in {\em EvoMUSART}. Springer, 2021.

\bibitem{trohidis2011multi}
Konstantinos Trohidis, Grigorios Tsoumakas, George Kalliris, and Ioannis
  Vlahavas,
\newblock ``Multi-label classification of music by emotion,''
\newblock {\em EURASIP}, vol. 2011, no. 1, pp. 1--9, 2011.

\bibitem{soleymani20131000}
Mohammad Soleymani, Micheal~N Caro, Erik~M Schmidt, Cheng-Ya Sha, and Yi-Hsuan
  Yang,
\newblock ``1000 songs for emotional analysis of music,''
\newblock in {\em Proc. of the 2nd ACM international workshop on Crowdsourcing
  for multimedia}, 2013.

\bibitem{lyberatos2023employing}
Vassilis Lyberatos, Spyridon Kantarelis, Eirini Kaldeli, Spyros Bekiaris,
  Panagiotis Tzortzis, Orfeas Menis-Mastromichalakis, and Giorgos Stamou,
\newblock ``Employing crowdsourcing for enriching a music knowledge base in
  higher education,''
\newblock in {\em AIET}. Springer, 2023, pp. 224--240.

\bibitem{chen2019harmony}
Tsung-Ping Chen and Li~Su,
\newblock ``Harmony transformer: Incorporating chord segmentation into harmony
  recognition,''
\newblock {\em neural networks}, 2019.

\bibitem{takeda2004rhythm}
Haruto Takeda, Takuya Nishimoto, and Shigeki Sagayama,
\newblock ``Rhythm and tempo recognition of music performance from a
  probabilistic approach,''
\newblock in {\em ISMIR}, 2004.

\bibitem{lisena2022midi2vec}
Pasquale Lisena, Albert Mero{\~n}o-Pe{\~n}uela, and Rapha{\"e}l Troncy,
\newblock ``Midi2vec: Learning midi embeddings for reliable prediction of
  symbolic music metadata,''
\newblock {\em Semantic Web}, vol. 13, no. 3, pp. 357--377, 2022.

\bibitem{pons2018end}
Jordi Pons~Puig, Oriol Nieto~Caballero, Matthew Prockup, Erik~M Schmidt,
  Andreas~F Ehmann, and Xavier Serra,
\newblock ``End-to-end learning for music audio tagging at scale,''
\newblock in {\em ISMIR}, 2018.

\bibitem{buisson2022ambiguity}
Morgan Buisson, Pablo Alonso-Jim{\'e}nez, and Dmitry Bogdanov,
\newblock ``Ambiguity modelling with label distribution learning for music
  classification,''
\newblock in {\em ICASSP}, 2022.

\bibitem{koops2019annotator}
Hendrik~Vincent Koops, W~Bas De~Haas, John~Ashley Burgoyne, Jeroen Bransen,
  Anna Kent-Muller, and Anja Volk,
\newblock ``Annotator subjectivity in harmony annotations of popular music,''
\newblock {\em Journal of New Music Research}, vol. 48, no. 3, pp. 232--252,
  2019.

\bibitem{Rudin2019}
Cynthia Rudin,
\newblock ``Stop explaining black box machine learning models for high stakes
  decisions and use interpretable models instead,''
\newblock {\em Nature Machine Intelligence}, vol. 1, no. 5, pp. 206--215, May
  2019.

\bibitem{holzapfel2018ethical}
Andre Holzapfel, Bob Sturm, and Mark Coeckelbergh,
\newblock ``Ethical dimensions of music information retrieval technology,''
\newblock {\em TISMIR}, vol. 1, no. 1, pp. 44--55, 2018.

\bibitem{mishra2017local}
Saumitra Mishra, Bob~L Sturm, and Simon Dixon,
\newblock ``Local interpretable model-agnostic explanations for music content
  analysis.,''
\newblock in {\em ISMIR}, 2017, vol.~53, pp. 537--543.

\bibitem{won2019toward}
Minz Won, Sanghyuk Chun, and Xavier Serra,
\newblock ``Toward interpretable music tagging with self-attention,''
\newblock {\em arXiv preprint arXiv:1906.04972}, 2019.

\bibitem{friberg2014using}
Anders Friberg, Erwin Schoonderwaldt, Anton Hedblad, Marco Fabiani, and Anders
  Elowsson,
\newblock ``Using listener-based perceptual features as intermediate
  representations in music information retrieval,''
\newblock {\em The Journal of the Acoustical Society of America}, vol. 136, no.
  4, pp. 1951--1963, 2014.

\bibitem{aljanaki2018data}
Anna Aljanaki and M.~Soleymani,
\newblock ``A data-driven approach to mid-level perceptual musical feature
  modeling,''
\newblock in {\em ISMIR}, 2018.

\bibitem{chowdhury2019towards}
Shreyan Chowdhury, Andreu~Vall Portabella, Verena Haunschmid, and Gerhard
  Widmer,
\newblock ``{Towards Explainable Music Emotion Recognition: The Route via
  Mid-level Features},''
\newblock 2019, ISMIR.

\bibitem{chowdhury2021towards}
Shreyan Chowdhury and Gerhard Widmer,
\newblock ``Towards explaining expressive qualities in piano recordings:
  Transfer of explanatory features via acoustic domain adaptation,''
\newblock in {\em ICASSP}. IEEE, 2021, pp. 561--565.

\bibitem{chowdhury2021tracing}
Shreyan Chowdhury, Verena Praher, and Gerhard Widmer,
\newblock ``Tracing back music emotion predictions to sound sources and
  intuitive perceptual qualities,''
\newblock {\em CoRR}, 2021.

\bibitem{gabrielsson2010role}
Alf Gabrielsson and Erik Lindstr{\"o}m,
\newblock ``The role of structure in the musical expression of emotions,''
\newblock {\em Handbook of music and emotion: Theory, research, applications},
  vol. 367400, pp. 367--44, 2010.

\bibitem{wedin1972multidimensional}
Lage Wedin,
\newblock ``A multidimensional study of perceptual-emotional qualities in
  music,''
\newblock {\em Scandinavian journal of psychology}, 1972.

\bibitem{han2022survey}
Donghong Han, Yanru Kong, Jiayi Han, and Guoren Wang,
\newblock ``A survey of music emotion recognition,''
\newblock {\em Frontiers of Computer Science}, vol. 16, no. 6, pp. 1--11, 2022.

\bibitem{SHAP}
Scott~M Lundberg and Su-In Lee,
\newblock ``A unified approach to interpreting model predictions,''
\newblock in {\em Advances in Neural Information Processing Systems 30},
  I.~Guyon, U.~V. Luxburg, S.~Bengio, H.~Wallach, R.~Fergus, S.~Vishwanathan,
  and R.~Garnett, Eds. Curran Associates, Inc., 2017.

\bibitem{xgboost}
Tianqi Chen and Carlos Guestrin,
\newblock ``Xgboost: A scalable tree boosting system,''
\newblock in {\em SIGKDD}. 2016, KDD '16, ACM.

\bibitem{mtg}
Dmitry Bogdanov, Minz Won, Philip Tovstogan, Alastair Porter, and Xavier Serra,
\newblock ``The mtg-jamendo dataset for automatic music tagging,''
\newblock in {\em Machine Learning for Music Discovery Workshop, ICML}, 2019.

\bibitem{tzanetakis2002musical}
George Tzanetakis and Perry Cook,
\newblock ``Musical genre classification of audio signals,''
\newblock {\em IEEE Transactions on speech and audio processing}, vol. 10, no.
  5, pp. 293--302, 2002.

\bibitem{kim2010music}
Youngmoo~E Kim, Erik~M Schmidt, Raymond Migneco, Brandon~G Morton, Patrick
  Richardson, Jeffrey Scott, Jacquelin~A Speck, and Douglas Turnbull,
\newblock ``Music emotion recognition: A state of the art review,''
\newblock in {\em ISMIR}, 2010.

\bibitem{wu2021omnizart}
Yu-Te Wu, Yin-Jyun Luo, Tsung-Ping Chen, I-Chieh Wei, Jui-Yang Hsu, Yi-Chin
  Chuang, and Li~Su,
\newblock ``Omnizart: A general toolbox for automatic music transcription,''
\newblock {\em Journal of Open Source Software}, vol. 6, no. 68, pp. 3391,
  2021.

\bibitem{KANTARELIS2023100754}
Spyridon Kantarelis, Edmund Dervakos, Natalia Kotsani, and Giorgos Stamou,
\newblock ``Functional harmony ontology: Musical harmony analysis with
  description logics,''
\newblock {\em Journal of Web Semantics}, vol. 75, pp. 100754, 2023.

\bibitem{bogdanov2013essentia}
Dmitry Bogdanov, Nicolas Wack, Emilia G{\'o}mez~Guti{\'e}rrez, Sankalp Gulati,
  Herrera Boyer, Oscar Mayor, Gerard Roma~Trepat, Justin Salamon,
  Jos{\'e}~Ricardo Zapata~Gonz{\'a}lez, Xavier Serra, et~al.,
\newblock ``Essentia: An audio analysis library for music information
  retrieval,''
\newblock ISMIR, 2013.

\bibitem{guidotti2018survey}
Riccardo Guidotti, Anna Monreale, Salvatore Ruggieri, Franco Turini, Fosca
  Giannotti, and Dino Pedreschi,
\newblock ``A survey of methods for explaining black box models,''
\newblock {\em ACM computing surveys (CSUR)}, vol. 51, no. 5, pp. 1--42, 2018.

\bibitem{tovstogan2021media}
Philip Tovstogan, Dmitry Bogdanov, and Alastair Porter,
\newblock ``Media-eval 2021: Emotion and theme recognition in music using
  jamendo,''
\newblock in {\em Proc. of the MediaEval 2021 Workshop, Online}, 2021, pp.
  13--15.

\bibitem{bour2021media}
Vincent Bour,
\newblock ``Frequency dependent convolutions for music tagging,''
\newblock in {\em Proc. of the MediaEval 2021 Workshop, Online}, 2021.

\bibitem{gtzan1}
Daisuke Niizumi, Daiki Takeuchi, Yasunori Ohishi, Noboru Harada, and Kunio
  Kashino,
\newblock ``Masked modeling duo: Learning representations by encouraging both
  networks to model the input,''
\newblock in {\em ICASSP}, 2023.

\bibitem{gtzan2}
Rodrigo Castellon, Chris Donahue, and Percy Liang,
\newblock ``Codified audio language modeling learns useful representations for
  music information retrieval,''
\newblock in {\em ISMIR}, 2021.

\bibitem{gomez2020joyful}
Juan~Sebasti{\'a}n G{\'o}mez~Ca{\~n}{\'o}n, Estefan{\'\i}a Cano, Herrera Boyer,
  Emilia G{\'o}mez~Guti{\'e}rrez, et~al.,
\newblock ``Joyful for you and tender for us: The influence of individual
  characteristics and language on emotion labeling and classification,''
\newblock ISMIR, 2020.

\bibitem{mckay2006musical}
Cory McKay and Ichiro Fujinaga,
\newblock ``Musical genre classification: Is it worth pursuing and how can it
  be improved?,''
\newblock in {\em ISMIR}, 2006, pp. 101--106.

\end{thebibliography}

\end{document}